\documentclass[showpacs,10pt,twocolumn,prl]{revtex4-1}

\usepackage{amsmath}
\usepackage{amssymb}
\usepackage{graphicx}
\usepackage{amssymb}
\usepackage{graphics}
\usepackage{epsfig}
\usepackage{color}
\usepackage{soul}

\begin{document}

\title{Structural and physical properties of the chiral antiferromagnet CeRhC$_2$}
\author{Yu Liu,$^{1}$ M. O. Ajeesh,$^{1}$ A. O. Scheie,$^{1}$ C. R. dela Cruz,$^{2}$ P. F. S. Rosa,$^{1}$ S. M. Thomas,$^{1}$ J. D. Thompson,$^{1}$ F. Ronning,$^{3}$ and E. D. Bauer$^{1}$}
\affiliation{$^{1}$MPA-Q, Los Alamos National Laboratory, Los Alamos, New Mexico 87545, USA\\
$^{2}$Neutron Scattering Division, Oak Ridge National Laboratory, Oak Ridge, Tennessee 37831, USA\\
$^{3}$Institute for Materials Science, Los Alamos National Laboratory, Los Alamos, New Mexico 87545, USA}
\date{\today}

\begin{abstract}
We report a study of the structural, magnetic, transport, and thermodynamic properties of polycrystalline samples of CeRhC$_2$. CeRhC$_2$ crystallizes in a tetragonal structure with space group $P4_1$ and it orders antiferromagnetically below $T_\textrm{N1} \approx$ 1.8~K. Powder neutron diffraction measurements reveal a chiral magnetic structure with a single propagation vector $Q_m = (1/2,1/2,0.228(5))$, indicating an antiferromagnetic arrangement of Ce magnetic moments in the $ab$-plane and incommensurate order along the $c$-axis with a root-mean-square ordered moment of $m_\textrm{ord}$= 0.68~$\mu_\textrm{B}$/Ce. Applying a magnetic field suppresses the N\'{e}el temperature $T_\textrm{N1}$ to zero near $\mu_0H_\textrm{c1}\sim0.75$ T. A second antiferromagnetic phase ($T_\textrm{N2}$), however, becomes apparent in electrical resistivity, Hall and heat capacity measurements in fields above 0.5~T and extrapolates to zero temperature at $\mu_0H_\textrm{c2}\sim1$ T. Electrical resistivity measurements reveal that LaRhC$_2$ is a semiconductor with a bandgap of $E_\textrm{g}\sim24$~meV; whereas, resistivity and Hall measurements indicate that CeRhC$_2$ is a semimetal with a low carrier concentration of $n\sim10^{20}$~cm$^{-3}$. With applied hydrostatic pressure, the zero-field antiferromagnetic transition of CeRhC$_2$ is slightly enhanced and CeRhC$_2$ becomes notably more metallic up to 1.36~GPa. The trend toward metallicity is in line with density-functional calculations that indicate that both LaRhC$_2$ and CeRhC$_2$ are semimetals, but the band overlap is larger for CeRhC$_2$, which has a smaller unit cell volume that its La counterpart. This suggests that the bandgap closes due to a lattice contraction when replacing La with Ce in RRhC$_2$ (R = rare-earth), in agreement with experimental results.
\end{abstract}
\maketitle

\section{INTRODUCTION}

Symmetry$-$ or the lack of it$-$ plays an important role in dictating the properties of materials \cite{Gross}. Chiral materials with crystal structures that do not have inversion, mirror, or roto-inversion symmetries often show exotic quantum phenomena \cite{Hasan}. For example, the chiral material CoSi, which crystallizes in the cubic $B20$ structure (space group $P2_13$),  hosts unconventional chiral fermions with a large topological charge that is connected by giant Fermi arcs \cite{CoSi1,CoSi2}. Likewise, isostructural MnSi exhibits an electrical magnetochiral effect (EMC), i.e., the non-reciprocal transport of conduction electrons depending on the inner product of current and magnetic field \cite{Tokura}. Moreover, broken inversion symmetry in chiral magnetic materials allows for the Dzyaloshinskii-Moriya (DM) interaction \cite{D,M}, which tends to induce canting of the spins away from a  collinear arrangement. The competition between the DM and other magnetic interactions that favor collinear spins may lead to modulated spin structures, such as chiral helimagnetic order, conical structures, chiral soliton lattices \cite{soli1,soli2,soli3}, and chiral magnetic skyrmion lattices \cite{skyr1,skyr2,skyr3}. To realize these emergent quantum states and to understand the underlying physics that dictates their properties, exploring new materials that might host these states is highly desired.

The ternary rare-earth carbides RTC$_2$ (R = rare-earth metal, T = transition metal) are interesting because they exhibit a diversity of inversion and/or time-reversal symmetry breaking phenomena \cite{Ray,Tsokol,Hoffmann,Matsuo,Bodak1,Bodak2,arXiv}, and are potential candidates for the above quantum states. The RNiC$_2$ (R=rare earth) compounds crystallize in the CeNiC$_2$-type orthorhombic structure with the $Amm2$ space group that lacks inversion symmetry. RNiC$_2$ materials have been widely studied and exhibit a variety of magnetic ground states \cite{Onodera,Kotsanidis,Jeitschko}, superconductivity (SC) \cite{Lee,Hillier,Hirose,Yanagisawa} that includes the possibility of spin-triplet superconductivity in  CeNiC$_2$ under pressure \cite{Katano}, and a complex interplay between these phases and charge density waves (CDW) \cite{Shimomura,Hanasaki1,Kolincio1,Hanasaki2,Kolincio2,Steiner,Kolincio}. CeNiC$_2$ is an example of the complex interplay of degrees of freedom in the RNiC$_2$ series. CeNiC$_2$ undergoes three successive magnetic phase transitions, from a paramagnetic (PM) to an incommensurate antiferromagnetic (AFM) phase at $\sim$ 20~K, a commensurate AFM phase below 10~K and finally to a ferromagnetic/ferrimagnetic state below $\sim$ 2~K \cite{Motoya,Pecharsky,Bhattacharyya}. RCoC$_2$ compounds adopt the same CeNiC$_2$-type structure for heavy rare-earth members (R = Gd-Lu) and exhibit ordering temperatures comparable to RNiC$_2$ with the same rare-earth metal \cite{Schafer1,Schafer2}. However, compounds with light rare-earth (R = La-Nd) members adopt the monoclinic CeCoC$_2$-type structure with the $C_1c_1$ space group, which is a distorted modification of the CeNiC$_2$-type structure involving  tilting of the C$_2$ pairs \cite{Jeitschko}. CeCoC$_2$  is a Kondo-lattice material with a characteristic Kondo temperature $T_\textrm{K} \sim$ 30~K \cite{Michor}.

Replacing Ni or Co with larger Rh produces RRhC$_2$, which crystallizes in a tetragonal CeRhC$_2$-type structure for R = La, Ce with the chiral $P4_1$ space group, but adopts the CeNiC$_2$ structure-type for R = Pr-Sm \cite{Tsokol}. CeRhC$_2$ forms in an interesting chiral structure with Ce atoms forming a helix along the $c$-axis. Aside from structural details, the physical properties of CeRhC$_2$ have not been investigated in detail, especially at low temperatures \cite{SCES}. In this paper, we report a detailed study on polycrystalline samples of CeRhC$_2$, along with its nonmagnetic analog LaRhC$_2$, by dc magnetic susceptibility, isothermal magnetization, heat capacity, longitudinal and Hall resistivity, and powder neutron diffraction measurements.

\section{EXPERIMENTAL DETAILS}

Polycrystalline samples of LaRhC$_2$ and CeRhC$_2$ were prepared by arc melting high purity La or Ce, Rh, and C mixtures with a total mass $\sim$ 1~g on a water-cooled copper hearth under an argon atmosphere with a Zr getter. A stoichiometric ratio of La:Rh:C=1:1:2 and a ratio of Ce:Rh:C = 1:0.9:2 or 1:1:2.4 produced the highest quality samples of LaRhC$_2$ and CeRhC$_2$, respectively, with the least amount of impurities. The samples were wrapped in tantalum foil, sealed in an evacuated silica tube, and then annealed for two weeks at 1000 $^\circ$C for LaRhC$_2$. An annealing temperature of 1100 $^\circ$C yielded higher quality CeRhC$_2$, and small single crystals ($\sim$50 microns) were obtained when annealed at 1200 $^\circ$C.

Powder X-ray diffraction (XRD) patterns were collected for two hours for each using a Malvern Panalytical Empyrean diffractometer in the Bragg-Brentano geometry using Cu K-$\alpha$ radiation. Single crystal XRD of CeRhC$_2$ was collected at room temperature on a $\sim$ 10-$\mu$m single crystal by a Bruker D8 Venture single-crystal X-ray diffractometer equipped with Mo radiation at room temperature. Magnetization measurements were performed in a Quantum Design Magnetic Property Measurement System (MPMS) from 1.8 to 350~K and magnetic fields up to 6.5~T. Specific heat measurements were performed using a Quantum Design Physical Property Measurement System (PPMS) from 0.35 to 20~K and magnetic fields up to 9~T that utilizes a quasi-adiabatic thermal relaxation method. Longitudinal resistivity and Hall resistivity measurements on a parallelopiped cut from an arc-melted button were carried in a PPMS using a standard four-probe configuration with an ac resistance bridge (Lake Shore, model 372). If the voltage contacts were misaligned, it was possible to determine the Hall resistivity by taking the difference of transverse resistivity measured at positive and negative fields, i.e., $\rho_{\textrm{xy}} = (\rho_{\textrm{H+}}-\rho_{\textrm{H-}})/2$, to eliminate the longitudinal resistivity contribution. Resistivity measurements under hydrostatic pressure were carried out using a double-layered piston-cylinder-type pressure cell with Daphne 7373 oil as the pressure-transmitting medium. The pressure inside the sample space was determined at low temperatures by the shift of the superconducting transition temperature of a piece of lead.

Neutron diffraction experiments were carried out at the HB2A powder diffractometer \cite{NPD1} at Oak Ridge National Laboratory$'$s HFIR. A 6~g powder sample of CeRhC$_2$ was mounted in an aluminum can with a copper lid, pressurized under 10 bar helium gas, and placed in a $^3$He refrigerator. The diffraction pattern with neutron wavelength $\lambda$ = 2.41~\AA~ was collected at 4~K and at 0.3~K for 16 hours at each temperature.

Density functional theory (DFT) calculations were performed using the Perdew, Burke, and Ernzerhof exchange correlation functional \cite{PBE} and included spin-orbit coupling without relativistic local orbitals through a second variational method using the WIEN2K code \cite{wien2k}. For CeRhC$_2$, the Ce f-electron was treated as an open core (localized) state.

\section{RESULTS AND DISCUSSION}

\begin{figure}
\centerline{\includegraphics[scale=1]{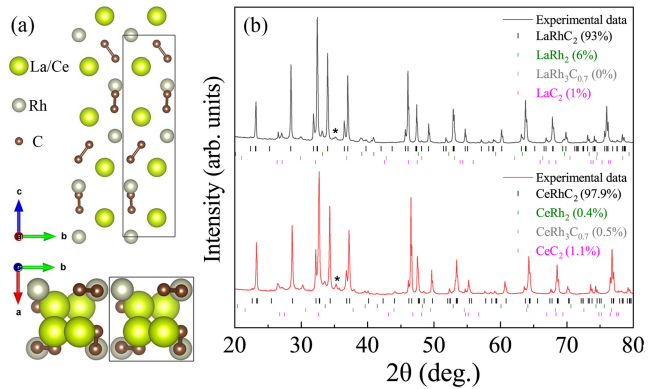}}
\caption{(Color online). (a) Crystal structure from both side and top views and (b) powder x-ray diffraction (XRD) patterns of LaRhC$_2$ and CeRhC$_2$. The peak marked with $*$ represents an unknown impurity phase.}
\label{XRD}
\end{figure}

The crystal structure of LaRhC$_2$ and CeRhC$_2$, first determined by Tsokol$'$ \emph{et al.} \cite{Tsokol}, is comprised of trigonal prisms of rare-earth atoms that are filled by Rh  atoms and C$_2$ pairs \cite{Hoffmann}. Compared to the CeNiC$_2$-type structure with two-dimensional (2D) C$_2$ sheets, LaRhC$_2$ and CeRhC$_2$ crystallize in a chiral structure with a more three-dimensional (3D) network resulting from tilting half of the C$_2$ pairs \cite{Momma}. Figure 1(a) depicts the chiral structure of LaRhC$_2$ and CeRhC$_2$ from both side and top views with a clear fourfold screw axis along the $c$ direction. Figure 1(b) shows the powder XRD patterns of LaRhC$_2$ and CeRhC$_2$; peaks can be well indexed in the $P4_1$ space group for the majority phase but a small amount of impurities ($\leq 7$\% total for LaRhC$_2$ and $\leq 3$\% total for CeRhC$_2$) also could be identified. Bragg peaks of impurities along with their volume percentages are included in Fig. 1(b). Rietveld refinement determined lattice parameters are $a$ = 3.970(1) {\AA} and $c$ = 15.331(1) {\AA} for LaRhC$_2$. The unit cell decreases to $a$ = 3.932(1) {\AA} and $c$ = 15.305(1) {\AA} for CeRhC$_2$, indicating a volume contraction $\sim$ 2.1\% relative to the La analog. Powder XRD parameters match well with $a$ = 3.93 {\AA} and $c$ = 15.29 {\AA} obtained from single crystal XRD of CeRhC$_2$ and are close to the published lattice parameters \cite{Hoffmann}.

\begin{figure}
\centerline{\includegraphics[scale=1]{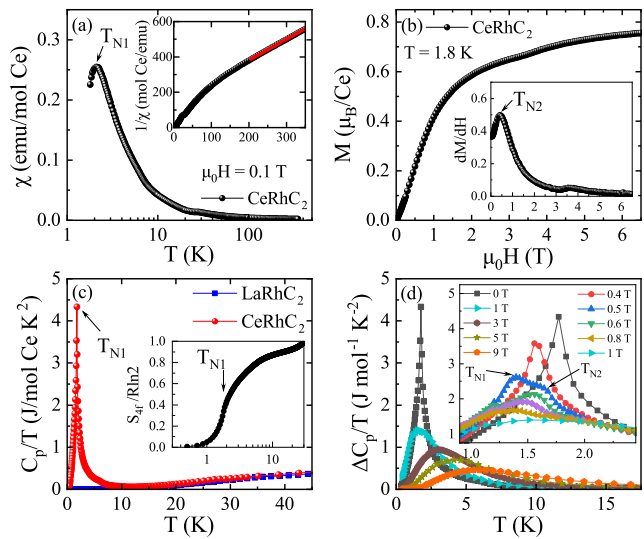}}
\caption{(Color online). (a) Temperature-dependent magnetic susceptibility $\chi$(T) of CeRhC$_2$ measured in a magnetic field of $\mu_0H$ = 0.1 T in zero-field-cooled (ZFC) mode. The inset of (a) shows the temperature-dependent inverse magnetic susceptibility $1/\chi$(T) and a linear fit from 200 to 350 K. (b) Field dependence of magnetization $M$($\mu_0$H) of CeRhC$_2$ at $T$ = 1.8~K. The inset plots the field derivative of $M$($\mu_0$H) and reveals a second magnetic transition at $T_\textrm{N2}$. A weak feature near 3.75~T is due to a second phase as discussed in the text. (c) Temperature dependence of specific heat $C_\textrm{p}/T$ of LaRhC$_2$ and CeRhC$_2$ in zero field. The inset of (c) shows the $4f$ contribution to the entropy $S_{4f}$(T)/Rln2 of CeRhC$_2$. (d) Temperature dependence of electronic specific heat $\Delta C_\textrm{p}/T$ in various magnetic fields.}
\label{MTH}
\end{figure}

Figure 2(a) shows the temperature dependence of the dc magnetic susceptibility $\chi$(T)=$M/\mu_0H$ measured in a field $\mu_0H$ = 0.1 T. A peak at $T_\textrm{N}$ $\approx$ 2.1~K is observed, in line with a previous report \cite{SCES}, reflecting an AFM transition. The temperature-dependent inverse susceptibility $1/\chi$(T) is plotted in the inset of Fig. 2(a). A linear fit from 200 to 350~K to a Curie-Weiss law, $\chi = C/(T-\theta)$, where $C$ is the Curie constant and $\theta$ is the paramagnetic Curie-Weiss temperature, yields $\theta$ = -121 K, but this does not reflect the sign of the magnetic exchange because of crystal field effects. The derived effective moment $\mu_{\textrm{eff}}$ = 2.60 $\mu_\textrm{B}/$Ce is close to the Ce$^{3+}$ free-ion moment of 2.54 $\mu_\textrm{B}$. Below about 150 K, $\chi$(T) deviates from Curie-Weiss behavior, most likely due to the effect of the crystalline electric field (CEF) splitting of the $J=5/2$ manifold. A crude estimate of the energy $\Delta_1$ of the first excited CEF doublet is estimated by fitting $\chi$(T) to $\chi = p_0 \chi_1 + p_{\Delta_1} \chi_2$, where $p_n$ is the Boltzmann population factor and $\chi_\textrm{n}$ is the Curie-Weiss susceptibility of the ground and excited multiplet. This function reproduces the temperature dependence of $\chi$(T) of CeRhC$_2$ (not shown) with $\Delta_1=27$~meV and high- (low-) temperature effective moment of $\mu_\textrm{eff}=2.4$~$\mu_\textrm{B}$ (1.7~$\mu_\textrm{B}$). Therefore, the ground state doublet is well separated from the first excited CEF doublet. Figure 2(b) displays the isothermal magnetization $M$(H) measured at $T$ = 1.8~K along with its first derivative $dM/dH$ in the inset. Above about 1~T, $M$(H) starts to saturate and reaches $\sim$ 0.75~$\mu_\textrm{B}$/Ce at 6.5~T. As discussed later, this high-field value of $M$ is close to the value of the ordered moment determined by neutron diffraction. $M$(H) does not follow a simple Brillouin function, which is reflected in a maximum in $dM/d\mu_0H$ near 0.4~T. On the basis of other results discussed below, this maximum is attributed to a field-induced transition to a different magnetic state at $T_\textrm{N2}$. Finally,  a weak anomaly occurs in $dM/dH$ near 3.75~T as well as in $\chi$(T) and $C_\textrm{p}/T$ [Fig. 2(c)] near $T \sim$ 30~K. These features are due to a small amount ($<$ 2\%) of  magnetic CeC$_2$ impurity \cite{Sakai1,Sakai2}, which orders antiferromagnetically at $T_\textrm{N}=30$~K and was also present in CeNiC$_2$ \cite{Pecharsky}. The magnetic entropy associated with the magnetic ordering of CeC$_2$ near $\sim$ 30~K is about 72 mJ/mol~K, suggesting $\sim$ 0.5 at.\% of CeC$_2$ in the CeRhC$_2$ sample.

\begin{figure}
\centerline{\includegraphics[scale=0.5]{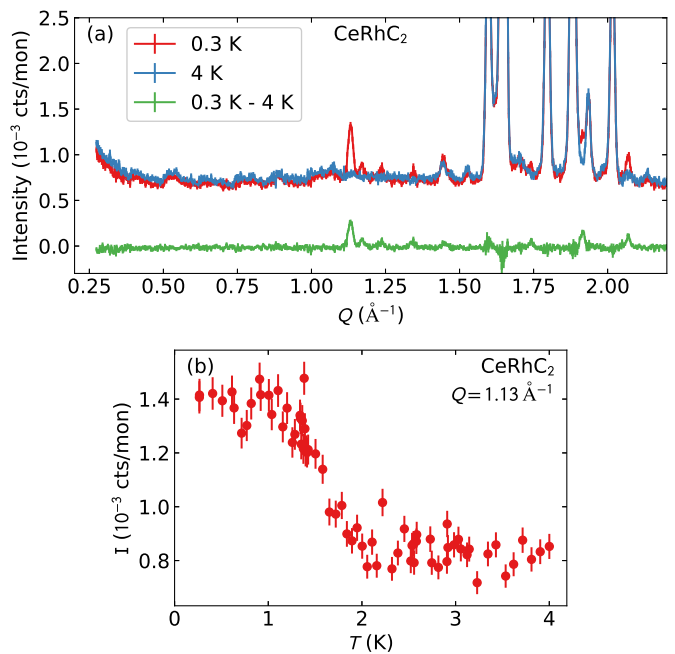}}
\caption{(Color online). (a) Powder neutron diffraction pattern of CeRhC$_2$ at 0.3 and 4~K, with the temperature-subtracted (0.3~K - 4~K) data shown in green below. Several (magnetic) Bragg peaks appear at low temperatures. (b) The temperature scan of the largest magnetic Bragg intensity at $Q=1.13$~\AA$^{-1}$, showing the onset of intensity is the same as the bulk anomalies.}
\label{NS}
\end{figure}

The specific heat divided by temperature, $C_\textrm{p}/T$ [Fig. 2(c)], shows a pronounced peak at $T_\textrm{N1}$ = 1.8~K, which is absent in the nonmagnetic, isostructural compound LaRhC$_2$. The magnetic entropy $S_{4f}$/Rln2 [Fig. 2(c) inset] is determined by integrating the $4f$ contribution to the specific heat $S_{4f} = \int{(\Delta C_\textrm{p}/T)}dT$, where $\Delta C_\textrm{p}$ is the $4f$ contribution obtained by subtracting the specific heat of LaRhC$_2$ from the measured specific heat of CeRhC$_2$. Only about 33\%$R$ln2 of entropy is released below $T_\textrm{N1}$. While the  Kondo effect could cause a significant reduction in the entropy released below $T_\textrm{N1}$, the Kondo effect appears to be weak in CeRhC$_2$, given that the $C_\textrm{p}/T$ as $T\rightarrow0$ in the AFM1 phase is small ($<$ 10~mJ/mol$^{-1}$K$^{-2}$ extrapolated from 0.35 K), and the carrier concentration is low (see below). This suggests that magnetic frustration may be responsible for the small amount of entropy released below $T_\textrm{N1}$. Figure 2(d) presents the temperature dependence of magnetic contribution $\Delta C_\textrm{p}/T$ of CeRhC$_2$ in various applied magnetic fields. The symmetric-like shape of $\Delta C_\textrm{p}/T$ at the AFM transition in zero field may indicate a first-order transition. To check for this possibility, a heat pulse applied just below the transition temperature to raise the temperature through the AFM transition revealed a small anomaly that may be consistent with latent heat from a first-order transition \cite{Allen}; however, no hysteresis was observed. We speculate that the transition may be weakly first-order, but further measurements are required to reach a firm conclusion regarding the nature of the transition. For $H<$ 1~T [Fig. 2(d) inset], $T_\textrm{N1}$ is gradually suppressed but a second anomaly, labelled $T_\textrm{N2}$, emerges in a field of 0.5~T. Both $T_\textrm{N1}$ and $T_\textrm{N2}$ move to lower temperatures with increasing field and become undetectable at fields of 1~T and greater. Instead of well-defined transitions, a broad maximum develops in $\Delta C_\textrm{p}/T$ around 1.6~K at $\mu_0H=$ 1~T, and it continues to shift to higher temperature and broaden with increasing field.

\begin{figure}
\centerline{\includegraphics[scale=1]{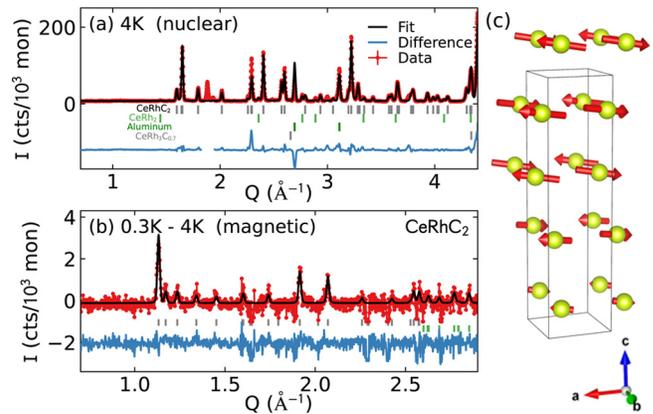}}
\caption{(Color online). Rietveld refinement of the nuclear (a) and magnetic (b) structure from powder neutron diffraction measurements of CeRhC$_2$. Panel (a) shows several peaks which have anomalous intensity, which could possibly be due to additional impurity phases. The temperature-subtracted data between 4 and 0.3~K in (b) was fit by the $\Gamma1$ irrep magnetic structure (see text), shown in panel (c).}
\label{NS_refinement}
\end{figure}

To shed light on the magnetic structure below $T_\textrm{N1}$ = 1.8~K in zero field, powder neutron diffraction data were collected at 0.3 and 4~K. As displayed in Fig. 3(a), new Bragg peaks appear below $T_\textrm{N1}$, which is illustrated by the difference curve of 0.3 and 4~K data. The intensity of the strongest magnetic Bragg peak versus temperature [Fig. 3(b)] shows an onset of $\sim$ 2~K, consistent with the AFM transition at $T_\textrm{N1}$ observed in bulk magnetic susceptibility and specific heat measurements. We therefore associate these temperature-dependent peaks with the onset of magnetic long-range order. The magnetic Bragg peaks can all be indexed by a single propagation vector $Q_m =(1/2, 1/2, 0.228(5))$, indicating commensurate antiferromagnetic order in the $ab$-plane, and incommensurate magnetic order along the $c$-axis.

\begin{figure}
\centerline{\includegraphics[scale=1]{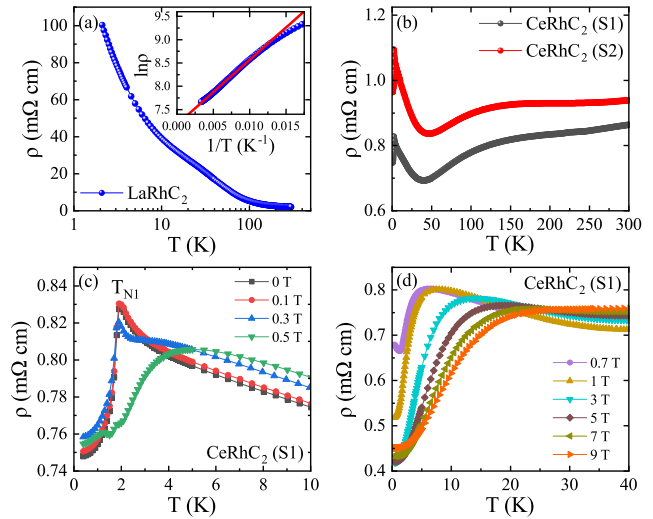}}
\caption{(Color online). Temperature dependence of electrical resistivity $\rho$(T) of (a) LaRhC$_2$ and (b) CeRhC$_2$ (two samples from the same batch are labeled as S1 and S2) in zero field. The inset in (a) shows an activated behavior,  ln$\rho$ versus $1/T$, along with a linear fit between 100-300~K. (c,d) Electrical resistivity $\rho$(T) for CeRhC$_2$ (S1) at low temperatures for several magnetic fields.}
\label{rho}
\end{figure}

To examine the magnetic structure in more detail, a Rietveld refinement was performed using the FULLPROF software package \cite{NPD2}, first to the 4~K nuclear scattering pattern to determine the normalization and resolution parameters, and then to fit the temperature-subtracted data as shown in Fig. 4. First, a magnetic refinement was carried out using an irreducible representation decomposition based on the derived propagation vector via the BasIreps program in FULLPROF. This approach yielded four possible irreducible representations, only one of which ($\Gamma1$) fit the observed diffraction pattern with a reduced $\chi^2 \approx 1$. However, given the sharpness of the heat capacity peak at $T_\textrm{N1}$, it is possible that the magnetic transition is first-order, in which case the magnetism need not be of a single irrep. In that case, there are a variety of statistically indistinguishable magnetic structures which fit the data, including spiral structures and uniaxial moment-modulated sinusoidal order (see Figs. 10 and 11 in APPENDIX A). Although the precise nature of the $c$-axis modulation can not be uniquely determined from the neutron powder diffraction data, the magnetic ground state clearly has: (i) commensurate AFM order in the $ab$-plane, (ii) (mostly) coplanar moments (a slight out-of-plane canting is possible), (iii) incommensurate magnetism along the $c$-axis, and (iv) a root-mean-square (RMS) ordered moment of $m_\textrm{ord}$=0.682(7)~$\mu_\textrm{B}$, close to the value obtained in the magnetization measurement [Fig. 2(b)]. These observations are true of all refined magnetic structures. Despite the ambiguity in the nature of the $c$-axis modulation, the CeRhC$_2$ magnetically ordered state at zero field is chiral, as the possible magnetic structures all lack inversion and mirror symmetry.

\begin{figure}
\centerline{\includegraphics[scale=1]{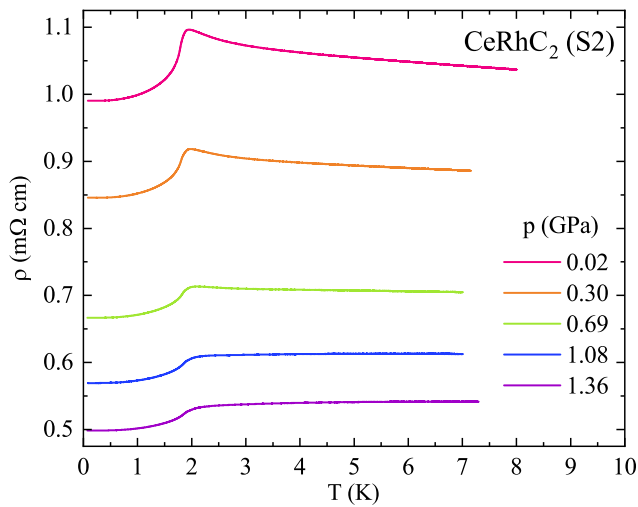}}
\caption{(Color online). Temperature dependence of electrical resistivity $\rho$(T) for CeRhC$_2$ (S2) under various pressures up to 1.36~GPa.(No offset applied to the data.)}
\label{rho_P}
\end{figure}

Having delineated the salient features of the complex magnetic structure of CeRhC$_2$, we now turn to an investigation of electrical transport properties. Figure 5(a) displays the temperature dependence of the electrical resistivity $\rho$(T) of LaRhC$_2$, which exhibits typical semiconducting behavior. The resistivity may be fit to an activated behavior, given by $\rho(T) = \rho_0 \textrm{exp}(E_\textrm{g}/k_\textrm{B}T)$, from 100 to 300~K [Fig. 5(a) inset], and the fit yields a small gap of $E_\textrm{g}\sim$ 24~meV, consistent with a previous result of $\approx$ 18 - 32~meV \cite{Hoffmann}. In contrast, samples of polycrystalline CeRhC$_2$ from the same batch show metallic behavior with similar features [Fig. 5(b)]; a broad hump between 100 and 150~K, a minimum around $T_{\textrm{min}}\sim$ 40~K, and a cusp at $T_\textrm{N1}$ = 1.9~K. It should be noted that the difference of absolute values of $\rho$(T) may be related to the relative amount of metallic impurities as stronger metallic behavior was observed (not shown) in samples with larger amounts of metallic impurities (e.g., CeC$_2$ \cite{CeC2}, CeRh$_2$ \cite{CeRh2}, CeRh$_3$C$_{0.7}$ \cite{CeRh3C}). With increasing magnetic field, $T_\textrm{N1}$ shifts to lower temperatures [Fig. 5(c)]. A second anomaly $T_\textrm{N2}$ emerges just above $T_\textrm{N1}\sim$ 1.25~K in a magnetic field of 0.5~T. The $T_\textrm{N1}$ transition disappears and an upturn appears at $T_\textrm{N2}$ $\approx$ 1.3~K in a magnetic field of 0.7~T [Fig. 5(d)], in agreement with specific heat measurements [Fig. 2(d) inset]. At low temperatures, a large negative magnetoresistance is observed, reaching about -48\% for higher fields ($>$ 1 T) at 2~K. The magnetoresistance becomes positive for temperatures above $\sim$ 20~K.

\begin{figure}
\centerline{\includegraphics[scale=0.9]{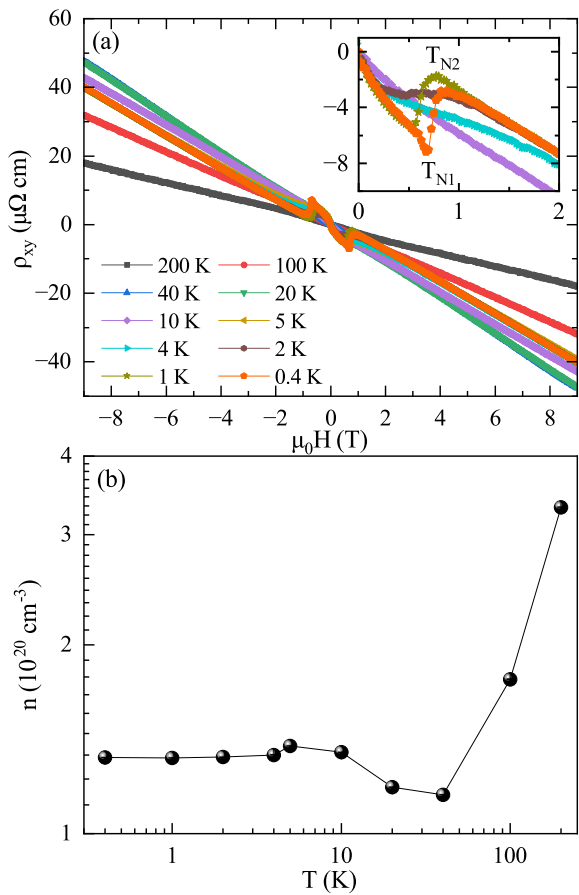}}
\caption{(Color online). (a) The field dependence of Hall resistivity $\rho_\textrm{xy}$($\mu_0$H) at various temperatures for CeRhC$_2$. The inset shows a magnification of the low field data. (b) The derived electron concentration $n$ by a linear fit to the $\rho_\textrm{xy}$($\mu_0$H) data from 3 to 9~T.}
\label{Hall}
\end{figure}

In the absence of evidence for a substantial Kondo effect in CeRhC$_2$, we would expect, and indeed find, a weak, positive response of magnetic order to applied pressure. Figure 6 shows the evolution of $\rho$(T) at low temperature with increasing hydrostatic pressures up to 1.36~GPa. At 0.02 GPa, the data resemble those at ambient pressure [Fig. 5(b)], exhibiting an AFM transition at $T_\textrm{N1}\approx$ 1.79~K as determined from the peak value of $d\rho$/$dT$ (see APPENDIX B). With increasing pressure, the transition broadens and  $T_\textrm{N1}$ increases linearly with a slope of 0.043 K/GPa, reaching $T_\textrm{N1}$ $\sim$ 1.85~K at 1.36~GPa. As shown in Fig. 12 in APPENDIX B, a weak shoulder in $d\rho$/$dT$ emerges just above the $T_\textrm{N1}$ at 1.08 and 1.36~GPa, pointing to a pressure-induced transition in CeRhC$_2$. Possibly, this second transition is related or identical to the field-induced transition $T_\textrm{N2}$, but this remains to be established. Nevertheless, what is rather remarkable about the data plotted in Fig. 6 is the overall decrease in the magnitude of $\rho$(T) with applied pressure, decreasing by about 0.5 m$\Omega$ cm at 1.36~GPa from its ambient-pressure value. This trend with modest applied pressure follows that in going from semiconducting LaRhC$_2$ to poorly metallic CeRhC$_2$ as the cell volume decreases by $\sim$ 2\%.

\begin{figure}
\centerline{\includegraphics[scale=1]{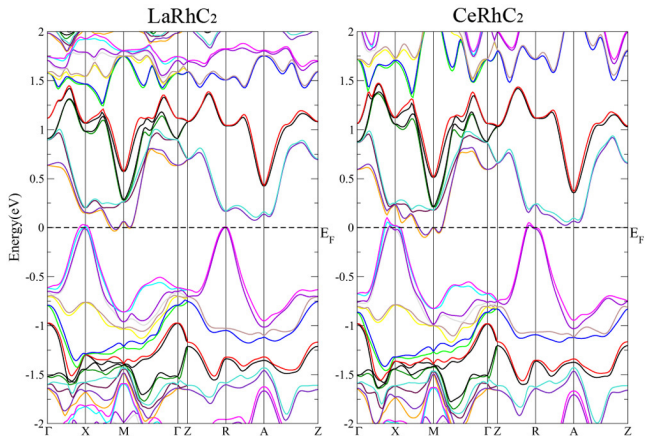}}
\caption{(Color online). Band structure calculated with spin-orbit coupling (SOC) for LaRhC$_2$ and CeRhC$_2$.}
\label{BandStruc}
\end{figure}

To determine the nature of transport carriers in CeRhC$_2$, the field dependence of Hall resistivity $\rho_{\textrm{xy}}$($\mu_0$H) was measured at various temperatures. $\rho_{\textrm{xy}}$ is almost linear with $\mu_0H$ at high temperatures, as shown in Fig. 7(a). The Hall coefficient ($R_\textrm{H}=\rho_{\textrm{xy}}$/$\mu_0H$) is negative, indicating dominant electron-like carriers in CeRhC$_2$. With decreasing temperature, $\rho_{\textrm{xy}}$ becomes nonlinear at low fields ($<$ 1~T), as magnified in the inset of Fig. 7(a), pointing to the likely multi-band nature of CeRhC$_2$. In the ordered state below 2~K, two sharp slope changes are clearly observed with a local minimum and maximum corresponding to $T_\textrm{N1}$ and $T_\textrm{N2}$, respectively. Assuming a simple single-band picture, a crude estimate of the carrier concentration $n$ from $R_\textrm{H}$ = $1/ne$, where $e$ is the electron charge, is obtained from a linear fit to $\rho_{\textrm{xy}}$ between 3 and 9~T. The so-derived carrier concentration as a function of temperature is plotted in Fig. 7(b), where $n$ decreases with decreasing temperature, reaches a minimum at $T$ = 40~K and shows a rather weak temperature dependence at lower temperatures. In spite of the over-simplification in estimating $n$, it seems likely that low values of carrier concentration are consistent with the semimetallic resistivity for CeRhC$_2$, which also features a local minimum around 40~K [Fig. 5(b)].

Transport properties of LaRhC$_2$ and CeRhC$_2$ are consistent with DFT band structure calculations of both compounds (Fig. 8). Due to the lack of inversion symmetry and the presence of spin-orbit coupling, all bands are singly degenerate except when they cross at time-reversal invariant momenta as well as the $k_z = \pi/c$ plane. Both compounds show a weak overlap between conduction and valence bands, which indicates compensated semimetallic behavior, with the band overlap notably stronger for CeRhC$_2$ than for LaRhC$_2$. Given that DFT calculations are known to overestimate bandwidths, it is anticipated that a reduction of the bandwidth would first open a small gap in LaRhC$_2$, consistent with the experimental observation that LaRhC$_2$ is semiconducting, while CeRhC$_2$ is a semimetal. Such corrections could be accomplished by including a modified Becke-Johnson (BJ) potential \cite{mBJ}, but are outside the scope of the present manuscript.

\begin{figure}
\centerline{\includegraphics[scale=1]{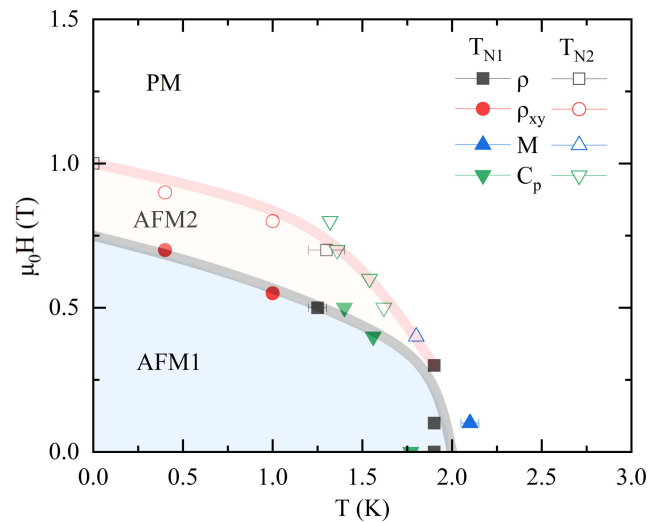}}
\caption{(Color online). The temperature-magnetic field ($T-\mu_0H$) phase diagram constructed from bulk properties for CeRhC$_2$. The PM and AFM represent paramagnetic and antiferromagnetic phases, respectively. The lines are guides to the eye.}
\label{phase_diagram}
\end{figure}

We summarize results of bulk properties measurements on CeRhC$_2$ in the temperature-magnetic field phase diagram given in Fig. 9. The solid and open symbols represent the transition into the AFM1 phase at $T_\textrm{N1}$ and the transition into the AFM2 phase at $T_\textrm{N2}$, respectively. Though neutron diffraction measurements reveal a chiral spin structure for AFM1, the nature of $T_\textrm{N2}$ remains unknown. Muon-spin rotation or neutron diffraction measurements in an applied field would provide useful information about the nature of the field-induced AFM2 phase. The growth and study of single crystalline CeRhC$_2$ will be useful for providing insight into the competition of magnetic interactions that leads to the stabilization of the chiral AFM1 phase and magnetic frustration in this interesting material.

\section{CONCLUSIONS}

Our study of magnetic, transport, and thermodynamic properties of polycrystalline samples of chiral CeRhC$_2$ show that it orders in a chiral magnetic structure in zero magnetic field and that another AFM phase appears at fields above $\sim$ 0.5 ~T. The nonmagnetic analog LaRhC$_2$ is semiconducting with a small bandgap that closes when replacing La with Ce. CeRhC$_2$ exhibits semi-metallic behavior with a low carrier concentration. Pressure appears to increase the band overlap in CeRhC$_2$ causing it to become more metallic up to 1.36~GPa, and reveals a possible second phase transition above $\sim1$~GPa. In future studies, it will be particularly interesting to search for exotic quantum states that arise from the chiral crystal and magnetic structures of CeRhC$_2$.

\section{ACKNOWLEDGMENTS}

Work at Los Alamos National Laboratory was performed under the auspices of the U.S. Department of Energy, Office of Basic Energy Sciences, Division of Materials Science and Engineering under project ``Quantum Fluctuations in Narrow-Band Systems". Y.L. and M.O.A. acknowledge a Director's Postdoctoral Fellowship through the Laboratory Directed Research and Development program. The diffraction experiment conducted at the High Flux Isotope Reactor of Oak Ridge National Laboratory was sponsored by the Scientific User Facilities Division, Office of Basic Energy Sciences, U.S. Department of Energy.

\newpage

\section{APPENDIX}
\subsection{A. REFINED MAGNETIC STRUCTURES}

\begin{figure}
\centerline{\includegraphics[scale=1]{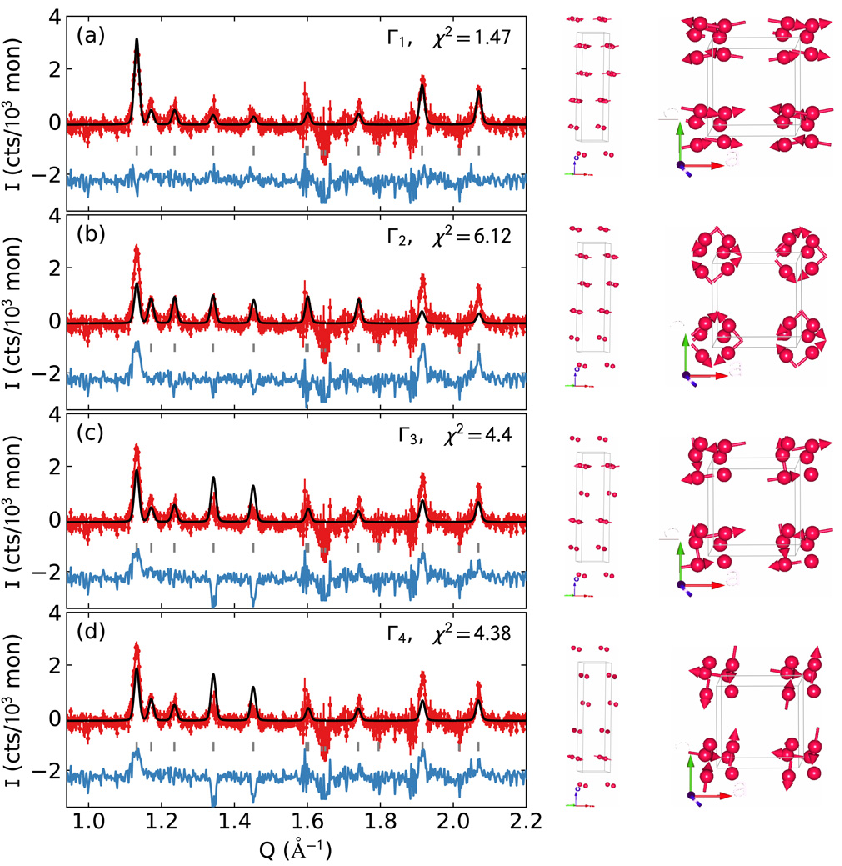}}
\caption{(Color online). Refined magnetic structures using the four irreducible representations of $P4_1$ with propagation vector $Q_m=(1/2, 1/2, 0.228(5))$. The best fits to the temperature subtracted data are shown on the left, and the refined structures (top and side views) are shown on the right. Clearly, the best fit is provided by $\Gamma1$.}
\label{NS_Gamma1}
\end{figure}

\begin{figure}
\centerline{\includegraphics[scale=1]{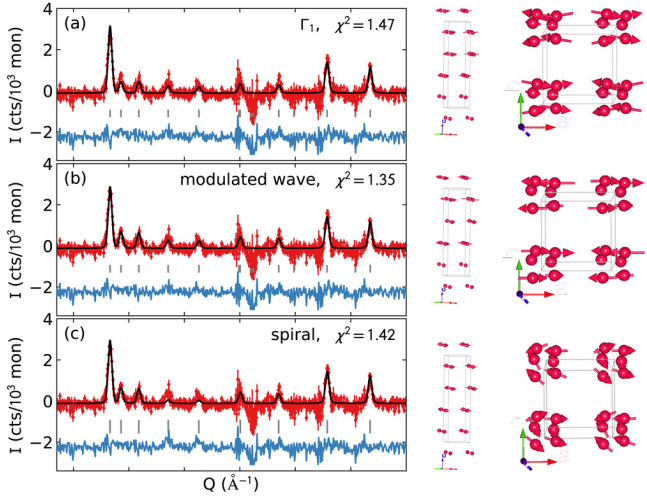}}
\caption{(Color online). Refined magnetic structures of arbitrary magnetic patterns, not necessarily those dictated by representation decomposition. Statistically equivalent (indeed, slightly better) fits are obtained by collinear sinusoidally modulated magnetism (b) and a coplanar spiral (c).}
\label{NS_other_structures}
\end{figure}

Figure 10 shows the refined magnetic structures of the four irreducible representations for the space group $P4_1$ and propagation vector $Q_m=(1/2, 1/2, 0.228(5))$. Reduced $\chi^2$ is noted in the upper right of each fit panel, where $\chi^2$ is calculated only in the vicinity of the magnetic peaks. Both visually and by the $\chi^2$ statistic, $\Gamma1$ provides the best fit to the magnetic diffraction.

If we lift the requirement for single irreducible representations, we can find many structures which fit the powder diffraction data. Some of these are shown in Figure 11, where the $\Gamma1$ structure (a) is compared against a collinear sinusoidal modulated structure (b) and a coplanar spiral of uniform moment size (c). By reduced $\chi^2$, these are indistinguishable in that the difference is less than 1. Therefore, we leave the precise details of the magnetic structure as an open question.

\subsection{B. TEMPERATUE DERIVATIVE OF RESISTIVITY UNDER PRESSURE}

\begin{figure}
\centerline{\includegraphics[scale=1]{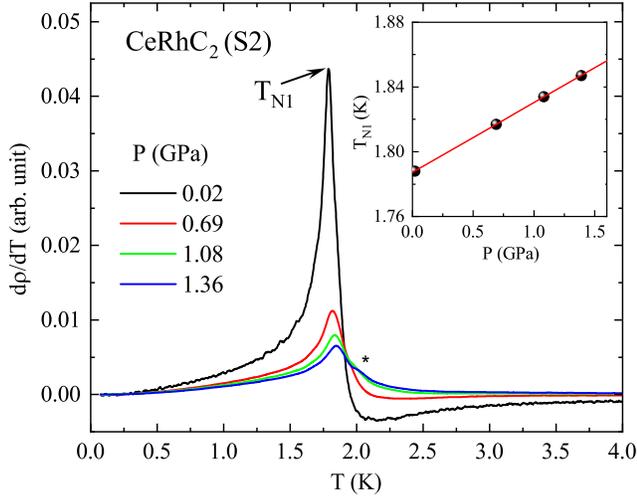}}
\caption{(Color online). Temperature dependence of temperature derivative electrical resistivity $d\rho/dT$ for CeRhC$_2$ (S2) under various pressures. Inset shows the evolution of $T_\textrm{N1}$ with pressure with a linear slope of $\sim$ 0.043 K/GPa.}
\label{drhodT}
\end{figure}

Figure 12 shows the temperature derivative electrical resistivity $d\rho/dT$ of CeRhC$_2$ (S2). The evolution of $T_\textrm{N1}$ with pressure is plotted in the inset of Fig. 12, showing a linear dependence with a slope of $\sim$ 0.043~K/GPa. Above $T_\textrm{N1}$, a weak anomaly emerges around 2~K as marked in asterisk for $P$ = 1.01 and 1.36 GPa, indicating the pressure-induced magnetic transition. The possibility of superconductivity may also be interesting to explore under high pressure.

\end{document}